\documentclass[showpacs,preprintnumbers]{revtex4}
\usepackage{amssymb}
\usepackage{graphicx}

\begin{document}

\title{Non-quasiparticle states in the core level spectra of ferromagnetic
semiconductors and half-metallic ferromagnets}
\author{V. Yu. Irkhin$^{1,2}$ and M. I. Katsnelson$^{2,3}$}
\address{$^{1}$Institute of Metal Physics, Ekaterinburg 620219, Russia\\
$^{2}$Department of Physics, Uppsala University, Uppsala 751 21,
Sweden\\
$^{3}$Department of Physics, University of Nijmegen, Nijmegen 6525 ED,
The Netherlands}

\begin{abstract}
The Green's functions that determine x-ray spectra are calculated in the $%
s-d $ exchange model of a saturated conducting ferromagnet in the presence
of the core hole. It is demonstrated that the core level (x-ray absorption,
emission and photoelectron) spectroscopy might be an efficient tool to
investigate the nonquasiparticle (NQP, spin-polaron) states in electron
energy spectrum since the core hole potential can enhance essentially their
spectral weight. NQP contributions to resonant x-ray scattering spectra can
be also much more pronounced than those to the the density of states.
\end{abstract}

\maketitle

\section{Introduction}

Connections between magnetic and electronic properties of materials are very
important both conceptually and from the point of view of possible
applications. Half-metallic ferromagnets (HMF) \cite{degroot,ufn,pickett},
ferromagnetic semiconductors \cite{nagaev,ohno,sarma}, and colossal
magnetoresistance materials \cite{nagaev1,edwreview,ziese} are of special
interest for spintronics (spin-dependent electronics) \cite{prinz}. Core
level spectroscopy techniques such as x-ray absorption, x-ray emission, and
photoelectron spectroscopies (xas, xes, and xps, correspondingly) give an
essential information about the electronic structure of these substances
(see, e.g., Refs.\onlinecite{yarm,yab,kurm,ola}). Unfortunately, a physical
background to understand corresponding experimental data is still very poor.
First, theoretical description of core-level spectra is a complicated
problem because of core hole effects \cite{Mah,noz}. Second, analysis of the
spectra is often based on the naive one-particle picture of density of
states (DOS), in particular, on the old Stoner picture of magnetism
formation. The latter theory is fully inapplicable for real itinerant
magnets where effects of electron correlations play a crucial role. These
effects are especially important in HMF which present a limiting case of
ultimately strong itinerant magnets \cite{ufn}.

Presence of the energy gap for the band states with one spin projection
results in a considerable contribution of the non-quasiparticle (NQP) states
which can occur in the gap \cite{ufn,edwards,IK1,IK2,Aus}. The formation of
NQP states is a consequence of formation of a spin polaron, i.e., a
superposition of minority-spin conduction electron states with the states of
majority-spin conduction electron \textit{plus} magnon excitations. This
effect has purely quantum nature (it disappears in the limit of large
localized spins) and essentially many-body character. In particular, the
singularities in the electron density of states owing to NQP contributions
are pinned to the Fermi level, similar to the standard Kondo effect.
Recently, a first-principle calculation of the NQP states was performed for
NiMnSb \cite{dmft}. The incoherent NQP states can exhibit themselves in the
tunneling phenomena, see Refs.\cite{transport,AItun,falko}. Moreover, it is
demonstrated in these papers that they determine $I-V$ characteristics of
tunnel junctions in the absence of normal quasiparticle transport, which is
the case in the presence of the energy gap. Therefore they are directly
related to spintronics. Besides that, the NQP states can be observed in the
photoelectron spectroscopy of conduction band \cite{IK1,Aus,KE} and in the
nuclear magnetic relaxation \cite{ufn,NMR}.

Effects of the NQP states in the core-level spectroscopy have been recently
discussed qualitatively for CrO$_2$ in Ref.\onlinecite{kurm}. Being almost
currentless, these states are in a sense strongly localized in the real
space. Therefore they yield a contribution to the elastic peak of x-ray
fluorescence, which is typical for insulators, but usually absent for
metals. In this work we present a quantitative microscopic theory which
gives a basis for the calculation of NQP contributions to various core level
spectra: xas, xes, xps and x-ray resonant scattering.

The corresponding two-particle Green's functions which determine the
spectral properties via Kubo formulas are treated in Sect. 2. It turns out
to be that in this case three-body problem (conduction electron, magnon, and
core hole) can be solved in a compact closed form suitable for calculations
with any bare density of electron states. To demonstrate qualitative
effects, we present in Sects. 3-5 the results of simple numerical
calculations of various core level spectra for the model semielliptic DOS.
We show that usually NQP contributions to core-level spectra are strongly
enhanced in comparison with those to initial DOS and therefore x-ray
spectroscopy might be a suitable tool to investigate them.

\section{Calculation of two-particle Green's functions}

To consider the core level problem we use the Hamiltonian of $s-d$ exchange
model in the presence of external potential $U$ induced by the core hole:

\begin{equation}
\mathcal{H}=\sum_{\mathbf{k}\sigma }t_{\mathbf{k}}c_{\mathbf{k}\sigma
}^{\dagger }c_{\mathbf{k}\sigma }+\varepsilon _{0}f^{\dagger }f-U\sum_{%
\mathbf{kk}^{\prime }\sigma }c_{\mathbf{k}\sigma }^{\dagger }c_{\mathbf{k}%
^{\prime }\sigma }f^{\dagger }f-I\sum_{\mathbf{qk}}\sum_{\alpha \beta }%
\mathbf{S_{q}}c_{\mathbf{k}\alpha }^{\dagger }\mbox {\boldmath $\sigma $}%
_{\alpha \beta }c_{\mathbf{k-q}\beta }+\sum_{\mathbf{q}}J_{\mathbf{q}}%
\mathbf{S}_{\mathbf{q}}\mathbf{S}_{-\mathbf{q}}  \label{H}
\end{equation}
Here $c_{\mathbf{k}\sigma }^{\dagger }$, $c_{\mathbf{k}\sigma }$ and $%
\mathbf{S}_{\mathbf{q}}$ are operators for conduction electrons and
localized spins in the quasimomentum representation, $f^{\dagger },f$ are
core hole creation and annihilation operators, $t_{\mathbf{k}}$ is the band
energy, $I$ is the parameter of the $s-d$ exchange interaction which is
assumed to be local, $\mathbf{\sigma }$ are the Pauli matrices, $J_{\mathbf{q%
}}$ are the Fourier transforms of the exchange parameters between localized
spins, which determine spin dynamics. The $s-d$ exchange model, where the
systems of local moments and current carriers are separated, describes a
ferromagnetic metal or degenerate semiconductor. Provided that the band
filling ($E_{F}$ calculated from the bottom) is smaller than the spin
splitting $\Delta =2|I|S$, this is the simplest model of HMF \cite{ufn}.

We use the method of double-time retarded Green's functions \cite{zubarev}.
The Green's function for operators $A$ and $B$
\begin{equation}
\langle \langle A|B\rangle \rangle _E^{\pm }=\int_{-\infty
}^0dt\,e^{iEt}\langle [e^{i\mathcal{H}t}A\,e^{-i\mathcal{H}t}\,,B]_{\pm
}\rangle ,\;\text{Im}E>0
\end{equation}
satisfies the equation of motion
\begin{equation}
E\langle \langle A|B\rangle \rangle _E^{\pm }=\langle [A,B]_{\pm }\rangle
+\langle \langle [A\,,\mathcal{H}]|B\rangle \rangle _E^{\pm }  \label{A}
\end{equation}
\begin{equation}
E\langle \langle A|B\rangle \rangle _E^{\pm }=\langle [A,B]_{\pm }\rangle
+\langle \langle A|[\mathcal{H},B\rangle \rangle _E^{\pm }  \label{B}
\end{equation}

We write down the equation of motion (\ref{A}) for the commutator
two-particle Green's function

\begin{equation}
G_{\mathbf{kk}^{\prime }}^\sigma (E)=\langle \langle c_{\mathbf{k}\sigma
}f|f^{\dagger }c_{\mathbf{k}^{\prime }\sigma }^{\dagger }\rangle \rangle _E
\end{equation}
which  takes into account core-level hole effects and determines x-ray
absorption and emission spectra \cite{Mah}. For the ferromagnetic state at
low temperatures (in the spin-wave region) we can pass to the magnon
representation for the spin operators, \
\begin{eqnarray*}
S_i^z &=&S-b_i^{\dagger }b_i, \\
S_i^{+} &=&(2S)^{1/2}b_i,S_i^{-}=(2S)^{1/2}b_i^{\dagger }
\end{eqnarray*}
Then we derive (the electron energy $E$ is referred to $\varepsilon _0$)
\begin{equation}
(E-t_{\mathbf{k}\sigma })G_{\mathbf{kk}^{\prime }}^\sigma (E)=(1-n_f-n_{%
\mathbf{k}}^\sigma )\left[ \delta _{\mathbf{kk}^{\prime }}-U\sum_{\mathbf{p}%
}G_{\mathbf{pk}^{\prime }}^\sigma (E)\right] -I\Phi _{\mathbf{k,k}^{\prime
}}^\sigma (E)  \label{g}
\end{equation}
where $t_{\mathbf{k}\sigma }=t_{\mathbf{k}}-\sigma I\langle S^z\rangle $ is
the Hartree-Fock spectrum$,n_{\mathbf{k}}^\sigma =n(t_{\mathbf{k}\sigma })$
is the Fermi function, $n_f$ is the occupation number for the $f$-hole in
the initial state, which is further on will be put to zero. We will take
into account the occupation numbers $n_{\mathbf{k}}^\sigma $ in a simple
ladder approximation which works well at small enough concentrations of
current carriers, except for the immediate vicinity of the Fermi edge. One
should note that the ladder approximation for the $s-d$ exchange model is
not the same as for the Hubbard model, but is much better owing to a proper
treatment of localized electron spins. For the Hubbard model without core
hole our approximation is equivalent to the Edwards-Hertz approach \cite
{edwards} which provides an adequate description of saturated ferromagnetic
state in a broad range of conduction electron concentrations. At the same
time, we do not treat here the problem of the x-ray edge singularity where
more advanced approaches are necessary \cite{Mah,noz}.

We use in Eq.(\ref{g}) the notation
\begin{eqnarray}
\Phi _{\mathbf{k-p,k}^{\prime }}^\sigma (E) &=&\sum_{\mathbf{r}}F_{\mathbf{%
k-p-r,r,k}^{\prime }}^\sigma (E) \\
F_{\mathbf{k-p,q,k}^{\prime }}^\sigma (E) &=&(2S)^{1/2}\langle \langle b_{%
\mathbf{q}}^\sigma c_{\mathbf{k}-\mathbf{p,}-\sigma }f|f^{\dagger }c_{%
\mathbf{k}^{\prime }\sigma }^{\dagger }\rangle \rangle _E, \\
b_{\mathbf{q}}^{+} &=&b_{-\mathbf{q}}^{\dagger },b_{\mathbf{q}}^{-}=b_{%
\mathbf{q}},  \nonumber
\end{eqnarray}
The Green's function $F$ satisfies the equation
\begin{eqnarray}
&&\ \ \ (E-t_{\mathbf{k-p,}-\sigma }+\sigma \omega _{\mathbf{q}})F_{\mathbf{%
k-p,q,k}^{\prime }}^\sigma (E)  \nonumber \\
\ &=&-U(1-n_{\mathbf{k-p}}^{-\sigma })\Psi _{\mathbf{q,k}^{\prime }}^\sigma
(E)-I(N_{\mathbf{q}}^\sigma +\sigma n_{\mathbf{k-p}}^{-\sigma })[2SG_{%
\mathbf{k-p+q,k}^{\prime }}^\sigma (E)+\sigma \Phi _{\mathbf{k-p+q,k}%
^{\prime }}^\sigma (E)]  \label{f}
\end{eqnarray}
where we have performed decouplings in spirit of the ladder approximation,
\[
\sum_{\mathbf{k}^{\prime \prime }}\langle \langle c_{\mathbf{k}^{\prime
\prime }-\sigma }^{\dagger }c_{\mathbf{k}^{\prime \prime }+\mathbf{q}\sigma
}c_{\mathbf{k}-\mathbf{p,}-\sigma }f|f^{\dagger }c_{\mathbf{k}^{\prime
}\sigma }^{\dagger }\rangle \rangle _E\rightarrow -n_{\mathbf{k-p}}^{-\sigma
}G_{\mathbf{k-p+q,k}^{\prime }}^\sigma (E),
\]
$\omega _{\mathbf{q}}$ is the magnon frequency, $\langle b_{-\mathbf{q}%
}^\sigma b_{\mathbf{q}}^{-\sigma }\rangle =N_{\mathbf{q}}^\sigma =\sigma
N(\sigma \omega _{\mathbf{q}}),N(\omega )=1/[\exp (\omega /T)-1]$ is the
Bose function,
\begin{equation}
\Psi _{\mathbf{q,k}^{\prime }}^\sigma (E)=\sum_{\mathbf{r}}F_{\mathbf{k-r,q,k%
}^{\prime }}^\sigma (E)
\end{equation}
For $U=0$ we have
\begin{equation}
G_{\mathbf{kk}^{\prime }}^\sigma (E)=(1-n_{\mathbf{k}}^\sigma )\delta _{%
\mathbf{kk}^{\prime }}G_{\mathbf{k}}^\sigma (E)
\end{equation}
where $G_{\mathbf{k}}^\sigma (E)$ is the one-electron Green's function of
the ideal crystal,
\begin{equation}
G_{\mathbf{k}}^\sigma (E)=\left[ E-t_{\mathbf{k}\sigma }-\Sigma _{\mathbf{k}%
}^\sigma (E)\right] ^{-1},\Sigma _{\mathbf{k}}^\sigma (E)=\frac{2I^2\langle
S^z\rangle Q_{\mathbf{k}}^\sigma }{1+\sigma IQ_{\mathbf{k}}^\sigma }
\label{g0}
\end{equation}
with
\[
Q_{\mathbf{k}}^{\uparrow }(E)=\sum_{\mathbf{q}}\frac{N_{\mathbf{q}}+n_{%
\mathbf{k+q}}^{\downarrow }}{E-t_{\mathbf{k+q\downarrow }}+\omega _{\mathbf{q%
}}},Q_{\mathbf{k}}^{\downarrow }(E)=\sum_{\mathbf{q}}\frac{1+N_{\mathbf{q}%
}-n_{\mathbf{k-q}}^{\uparrow }}{E-t_{\mathbf{k-q\uparrow }}-\omega _{\mathbf{%
q}}}
\]
Note that Eq.(\ref{g0}) \ yields correctly the exact Green's function in the
limit of an empty conduction band at $T=0$ \cite{Izyum,IK84,Aus}. The result
(\ref{g0}) corresponds to the approximation \cite{edwards} in the theory of
strong itinerant ferromagnetism (which is just the case of half-metallic
ferromagnets, cf.\cite{IK2}). This approximation can be justified with the
use of a Ward identity. Such an approximation in the $s-d$ exchange model is
also widely used for metals \cite{iks}.

For $I=0$ one obtains
\begin{eqnarray}
G_{\mathbf{kk}^{\prime }}^\sigma (E) &=&\frac{1-n_{\mathbf{k}}^\sigma }{E-t_{%
\mathbf{k}\sigma }}\left[ \delta _{\mathbf{kk}^{\prime }}-\frac{UP^\sigma (E)%
}{1+UP^\sigma (E)}\right]  \label{gu} \\
P^\sigma (E) &=&\sum_{\mathbf{k}}\frac{1-n_{\mathbf{k}}^\sigma }{E-t_{%
\mathbf{k}\sigma }}
\end{eqnarray}

In the general case, we have a three-particle problem (conduction electron,
core hole and magnon) which requires a careful mathematical investigation.
However, we can use the fact that the magnon frequencies are much smaller
than typical electron energies and, last but not least, that the resolution
of xas and xes methods is not sufficient to probe the energy scale of a
typical magnon frequency so we can put the latter to zero. Neglecting spin
dynamics the equations (\ref{g}), (\ref{f}) can be solved exactly in a
rather simple way provided that we consider the case of zero temperatures ($%
N_{\mathbf{q}}^{+}=0,N_{\mathbf{q}}^{-}=1$). Under these conditions $Q$ does
not depend on quasimomenta, and the scattering by the core hole in the
presence of a magnon is described by the same $\mathbf{q}$-independent
resolvent $P$. We derive from Eq.(\ref{f})
\begin{equation}
\Phi _{\mathbf{k-p+q,k}^{\prime }}^\sigma (E)=-\frac 1{1+\sigma IQ^\sigma
(E)}\left[ 2ISQ^\sigma (E)G_{\mathbf{k-p+q,k}^{\prime }}^\sigma
(E)+UP^{-\sigma }(E)\Psi _{\mathbf{q,k}^{\prime }}^\sigma (E)\right]
\label{eqfi}
\end{equation}
On summing over $\mathbf{p}$ we see that $\Psi _{\mathbf{q,k}^{\prime
}}^\sigma $ does nod depend on \textbf{q,}
\begin{equation}
\Psi _{\mathbf{q,k}^{\prime }}^\sigma (E)=\Psi _{\mathbf{k}^{\prime
}}^\sigma (E)=(2S)^{1/2}\langle \langle b^\sigma c_{-\sigma }f|f^{\dagger
}c_{\mathbf{k}^{\prime }\sigma }^{\dagger }\rangle \rangle _E=\sum_{\mathbf{%
pq}}F_{\mathbf{k-p,q,k}^{\prime }}^\sigma (E)=\sum_{\mathbf{p}}\Phi _{%
\mathbf{k-p,k}^{\prime }}^\sigma (E)
\end{equation}
(physically it means that the electron and magnon operator should belong to
the same perturbed site). Then the equation (\ref{eqfi}) can be solved in
terms of $\Psi $ to obtain
\begin{eqnarray}
\Psi _{\mathbf{k}^{\prime }}^\sigma (E) &=&-\frac{2ISQ^\sigma (E)}{%
1+UP^{-\sigma }(E)+\sigma IQ^\sigma (E)}R_{\mathbf{k}^{\prime }}^\sigma (E),
\label{psi} \\
R_{\mathbf{k}^{\prime }}^\sigma (E) &=&\sum_{\mathbf{k}}G_{\mathbf{kk}%
^{\prime }}^\sigma (E).
\end{eqnarray}
After substituting Eq.(\ref{psi}) into Eq.(\ref{eqfi}), summing over $%
\mathbf{p=q}$ and substituting the result into Eq.(\ref{g}) we obtain the
closed equation for the Green's function $G$

\[
\left[ E-t_{\mathbf{k}\sigma }-\Sigma ^\sigma (E)\right] G_{\mathbf{kk}%
^{\prime }}^\sigma (E)=(1-n_{\mathbf{k}}^\sigma )\left[ \delta _{\mathbf{kk}%
^{\prime }}-U\sum_{\mathbf{p}}G_{\mathbf{pk}^{\prime }}^\sigma (E)\right] -%
\frac{U\Sigma ^\sigma (E)P^{-\sigma }(E)}{1+UP^{-\sigma }(E)+\sigma
IQ^\sigma (E)}\sum_{\mathbf{p}}G_{\mathbf{pk}^{\prime }}^\sigma (E)
\]
Neglecting the factors $(1-n_{\mathbf{k}}^\sigma )$ which is possible at
small band filling (anyway, they are irrelevant for our purposes), we have
the standard result for the impurity scattering (\ref{gu}) with the
renormalized energy spectrum $E_{\mathbf{k}\sigma }=t_{\mathbf{k}\sigma
}+\Sigma ^\sigma (E)$ and the effective impurity potential
\begin{equation}
U_{ef}^\sigma (E)=U\left[ 1+\frac{\Sigma ^\sigma (E)P^{-\sigma }(E)}{%
1+UP^{-\sigma }(E)+\sigma IQ^\sigma (E)}\right]  \label{uef}
\end{equation}

Then the local density of states is given by
\begin{equation}
N_{\mathrm{loc}}^{\sigma }(E)=-\frac{1}{\pi }\mathrm{Im}G_{00}^{\sigma }(E)
\label{gsp}
\end{equation}
with
\begin{equation}
G_{00}^{\sigma }(E)=\sum_{\mathbf{kk}^{\prime }}G_{\mathbf{kk}^{\prime
}}^{\sigma }(E)=R_{\sigma }(E)+T^{\sigma }(E)R_{\sigma }^{2}(E)=\frac{%
R_{\sigma }(E)}{1+U_{ef}^{\sigma }(E)R_{\sigma }(E)}  \label{g00}
\end{equation}
where
\begin{equation}
R_{\sigma }(E)=G_{00}^{(0)\sigma }(E)=\sum_{\mathbf{k}}G_{\mathbf{k}%
}^{\sigma }(E),
\end{equation}
$G_{\mathbf{k}}^{\sigma }(E)$ is given by Eq.(\ref{g0}), and the $T$-matrix
has the form
\begin{equation}
T^{\sigma }(E)=-\frac{U_{ef}^{\sigma }(E)}{1+U_{ef}^{\sigma }(E)R_{\sigma
}(E)}  \label{T}
\end{equation}

\section{X-ray absorption and emission spectra}

Generally speaking, theoretical investigation of core level spectra requires
a numerical calculations with account of realistic bandstructure. We
restrict ourselves to some model examples. The resolvents in the complex
plane can be calculated analytically for the simple semielliptical bare
density of states (see Appendix).

The picture of the NQP contributions to the DOS of an ideal crystal is
described by the Green's function (\ref{g0}). This was considered in Refs. %
\onlinecite{Aus,IK1} for a degenerate ferromagnetic semiconductor and
discussed in detail for a half-metallic ferromagnet \cite{ufn}. Remember
that the energy gap occurs in the case where $\Delta =2|I|S>E_F.$ We have

\begin{equation}
\delta N^\sigma (E)=-\frac 1\pi \mathrm{Im}R_\sigma (E)=-\frac 1\pi \mathrm{%
Im}{\normalsize \Sigma }(E)\left| R_\sigma ^{\prime }(E)\right|  \label{bare}
\end{equation}

As demonstrates an analysis of the electron-magnon interaction, the picture
turns out to be different for two possible signs of the $s-d$ exchange
parameter $I$. For $I<0$ spin-up NQP states are present below the Fermi
level as an isolated region (Fig.1): occupied states with the total spin $%
S-1/2$ are a superposition of the states $|S\rangle |\downarrow \rangle $
and $|S-1\rangle |\uparrow \rangle $. On the contrary, for $I>0$ the
spin-down NQP scattering states form a ``tail'' of the upper spin-down band,
which starts from $E_F$ (Fig.2) since the Pauli principle prevents the
electron scattering into occupied states. Of course, the jumps at the Fermi
level are in fact smeared at the scale of the magnon energies $\overline{%
\omega }$.

Since xas probes empty states and xes occupied states, the quantity (\ref
{gsp}) describes the absorption spectrum for $E>E_F$ \ and emission spectrum
for $E<E_F.$ \ As follows from above results, the picture observed in the
core level spectroscopy is determined by more complicated integral equations
in comparison with the DOS problem. Therefore some new circumstances occur
for the NQP contributions. The numerical results in our simple model are
shown in Figs.3-4, the local density of states for $U=0$ (which coincides
with usual DOS in that case) being also presented for comparison. To take
into account core level broadening, we introduce in the quantity (\ref{gsp})
a finite damping $\delta $ near the Fermi level (see Appendix).

For $I>0$ the results (\ref{uef})-(\ref{g00}) provide full solution of the
Kondo problem for an impurity in the ferromagnet, corresponding to the
parquet approximation \cite{Abr}. On the other hand, for $I<0$ the situation
is complicated by the presence of the ``false'' Kondo divergence in the
quantity (\ref{T}), similar to the same approximation in the standard Kondo
problem \cite{Suhl}. Formally, the \textit{T}-matrix has a pole, and DOS
above the Fermi level even turns out to be negative. Generally speaking,
more advanced approximations are needed to solve this problem. However, this
difficulty is not important for the x-ray problem where a large damping is
always present, and experiments are performed at sufficiently high
temperatures with rather poor resolution (as compared with a scale of the
``Kondo temperature''). The case of very low temperatures and small damping
requires a special treatment which will be given elsewhere.

To leading order in $U$ and $I$ we obtain
\begin{eqnarray}
\delta N_{\mathrm{loc}}^\sigma (E) &=&\frac{1-\mathop{\rm Re}(U_{ef}^{\sigma
}(E)/\Sigma ^{\sigma }(E))\left| R_\sigma ^2(E)/R_\sigma ^{\prime
}(E)\right| }{\left| 1+U_{ef}^\sigma (E)R_\sigma (E)\right| ^2}\delta
N^\sigma (E)  \nonumber \\
&&-\frac{\mathop{\rm Re}(U_{ef}^{\sigma }(E)/P^{-\sigma }(E))\left| R_\sigma
(E)\right| ^2}{\left| 1+U_{ef}^\sigma (E)R_\sigma (E)\right| ^2}\frac 1\pi
\mathrm{Im}P^{-\sigma }(E)  \label{nn}
\end{eqnarray}
The term in Eq.(\ref{nn}) with $\mathrm{Im}P^{-\sigma }(E)$ results in a
smooth contribution to the spectrum. In particular, it is non-zero in the
energy gap. Note that for the emission spectra such a term is absent. The
NQP contributions to the absorption \ and emission spectra, that are
proportional to $\delta N^\sigma (E),$ occur for $I>0$ and $I<0$ only,
respectively.

The NQP contribution is the only minority contribution to $\delta N^\sigma
(E)$ near the Fermi level. As for majority contribution, this is smooth at
the Fermi level and can be roughly treated in the Hartree-Fock approximation
and obtained by a shift of the quasiparticle minority contribution by the
spin splitting $\Delta $ (Figs. 1,2).

One can see from Fig.3 that the upturn of the NQP tail which occurs for $I>0$
becomes somewhat more sharp, although the jump near $E_F$ weakens. For $I<0$
the spectral weight of NQP contributions increases in the presence of the
core hole too (Fig.4)$.$ These effects have a simple physical interpretation.

Since $U_{ef}^\sigma (E)>0$ and for small band filling $R_\sigma (E)<0$ near
$E_F,$ the denominator of the expression (\ref{nn}) results in a
considerable enhancement of NQP contributions to the spectra in comparison
with those to DOS. However, effects of interaction $U$ do not reduce to a
constant factor in the self-energy, but turn out to be non-trivial. Strong
interaction with the core hole results in a deformation of conduction band.
With increasing $U$ the spectral density passes to the band bottom. This
effect is especially important for the NQP states since they lie in this
region. Therefore the spectral weight of the NQP states increases. However,
with further increasing $U$ (at very large, in fact unrealistic values) a
bound state is formed near the band bottom, and the NQP spectral weight
becomes suppressed owing to factor of $U$ in the denominator of the
expression (\ref{uef}).

The temperature dependence of the spectra at relatively high temperatures
can be roughly taken into account by neglecting short-range order, i.e.,
formally, by neglecting $\mathbf{q}$-dependence of spin correlation
function. We have just to replace $N_{\mathbf{q}}\rightarrow S-\overline{S}$
in the resolvent $Q,$ so that effects of finite temperatures result in a
decrease of the jumps and in an increase of NQP density of states in the
energy gap.

\section{Photoelectron spectroscopy}

To discuss xps we consider the anticommutator core level Green's function
\begin{equation}
G_{f}^{{}}(E)=\langle \langle f|f^{\dagger }\rangle \rangle _{E}=\left[
E-\varepsilon _{0}-\Sigma _{f}(E)\right] ^{-1}  \label{ff}
\end{equation}
By using both the equations (\ref{A}) and (\ref{B}) the self-energy $\Sigma
_{f}(E)$ is expressed in terms of an irreducible Green's function,
\begin{equation}
\Sigma _{f}(E)=U^{2}\sum_{\sigma \sigma ^{\prime }}\langle \langle c_{\sigma
}^{\dagger }c_{\sigma }f|f^{\dagger }c_{\sigma ^{\prime }}^{\dagger
}c_{\sigma ^{\prime }}\rangle \rangle _{E}^{\mathrm{irr}}  \label{sff}
\end{equation}
(the superscript ``irr'' means that the contributions divergent as $%
(E-\varepsilon _{0})^{-n}$ should be omitted in the equations of motion for
the Green's function (\ref{sff}), cf. Ref.\onlinecite{Aus}).

To evaluate the Green's function (\ref{sff}) we pass to the time
representation and use the simplest decoupling (the vertices are neglected),
\begin{equation}
\Sigma _f(E)=-U^2\sum_\sigma \int_0^\infty dte^{iEt}\langle c_\sigma
^{\dagger }(t)c_\sigma \rangle \langle c_\sigma (t)f(t)f^{\dagger }c_\sigma
^{\dagger }\rangle
\end{equation}
Using the spectral representation for the correlation functions of $c$%
-operators we obtain
\begin{equation}
\Sigma _f(E)=\frac{U^2}\pi \sum_\sigma \int dE^{\prime }f(E^{\prime })%
\mathrm{Im}G_{00}^{(0)\sigma }(E^{\prime })G_{00}^\sigma (E^{\prime }-E)
\end{equation}
Thus the photoelectron spectrum contains NQP contributions both from initial
and final states, the hole effects being important only for final states.
Photocurrent can be in principle resolved in spin projection of
photoelectrons, which makes possible to observe the NQP contributions.

Since only the states below $E_{F}$ are observed in xps, we have to treat
spin up NQP states for $I<0$. The leading quasiparticle contribution from
spin down states reads
\begin{equation}
\delta \Sigma _{f\downarrow }(E)\simeq U^{2}\sum_{\mathbf{kk}^{\prime }}%
\frac{n_{\mathbf{k\downarrow }}(1-n_{\mathbf{k}^{\prime }\mathbf{\downarrow }%
})}{E-t_{\mathbf{k\downarrow }}+t_{\mathbf{k}^{\prime }\mathbf{\downarrow }}}%
=U^{2}\Pi _{0}(E)
\end{equation}
Despite of absence of spin up band states (at least, in our simple model of
HMF), transitions in NQP states with $\sigma =\uparrow $ (see Fig.2) take
place owing to electron-magnon scattering. The corresponding NQP
contribution has the structure
\begin{equation}
\delta \Sigma _{f\uparrow }(E)=\delta \Sigma _{f\uparrow }^{(1)}(E)+\delta
\Sigma _{f\uparrow }^{(2)}(E)
\end{equation}
with
\begin{eqnarray}
\delta \Sigma _{f\uparrow }^{(1)}(E) &\propto &-2I^{2}SU^{2}R_{\uparrow
}^{\prime }(E)\Pi _{1}(E), \\
\Pi _{1}(E) &=&\sum_{\mathbf{kk}^{\prime }}\frac{n_{\mathbf{k\downarrow }}}{%
E-t_{\mathbf{k\downarrow }}+t_{\mathbf{k}^{\prime }\mathbf{\uparrow }}} \\
\delta \Sigma _{f\uparrow }^{(2)}(E) &\propto &(2I^{2}S)^{2}U^{2}\left[
R_{\uparrow }^{\prime }(E)\right] ^{2}\Pi _{2}(E), \\
\Pi _{2}(E) &=&\sum_{\mathbf{kk}^{\prime }}\frac{n_{\mathbf{k\downarrow }}n_{%
\mathbf{k}^{\prime }\mathbf{\downarrow }}}{E-t_{\mathbf{k\downarrow }}+t_{%
\mathbf{k}^{\prime }\mathbf{\downarrow }}}
\end{eqnarray}

To leading order in $I$ only initial NQP states make a contribution. The
term $\delta \Sigma _f^{(1)}(E)$ has a threshold corresponding to the energy
gap, but is smeared practically over the whole band (Fig.5). The high-order
contribution $\delta \Sigma _f^{(2)}(E)$ starts from small $E$ and is more
singular,
\begin{eqnarray}
\mathrm{Re}\delta \Sigma _f^{(2)}(E) &\propto &E\ln |E|, \\
\mathrm{Im}\delta \Sigma _f^{(2)}(E) &\propto &E|E|
\end{eqnarray}

\section{Resonant x-ray scattering}

Now we consider NQP effects in resonant x-ray scattering processes. It was
observed recently \cite{kurm} that the elastic peak of the x-ray scattering
in CrO$_2$ is observed which is more pronounced than in usual Cr compounds,
e.g., the elemental chromium. The authors of this work have put forward some
qualitative arguments that the NQP states may give larger contributions to
resonant x-ray scattering than usual itinerant electron states. Here we
shall treat this question quantitatively and estimate explicitly the
corresponding enhancement factor. The intensity of resonant x-ray emission
induced by the photon with the energy $\omega $ and polarization \thinspace $%
q$ is given by the Kramers-Heisenberg formula \cite{sakurai,yab,gelm}
\begin{equation}
\mathcal{I}_{q^{\prime }q}(\omega ^{\prime },\omega )\propto \sum_n\left|
\sum_l\frac{\langle n|C_{q^{\prime }}|l\rangle \langle l|C_q|0\rangle }{%
E_0-\omega ^{\prime }-E_l-i\Gamma _l}\right| ^2\delta (E_n+\omega ^{\prime
}-E_0-\omega )  \label{xray}
\end{equation}
Here $q^{\prime },\omega ^{\prime }$ are the polarization and energy of the
emitted photon, $|n\rangle ,|0\rangle $ and $|l\rangle $ are the final,
initial and intermediate states of the scattering system, respectively, $E_i$
are the corresponding energies, $C_q$ is the operator of the dipole moment
for the transition, which is proportional to $fc+c^{\dagger }f^{\dagger }$.
For simplicity we will assume hereafter that $\Gamma _l$ does not depend on
the intermediate state, $\Gamma _l=\Gamma $, and take into account only the
main x-ray scattering channel where the hole is filled from the conduction
band (for a more general multichannel consideration, see, e.g., Ref.%
\onlinecite{sokolov}). Assuming also that the electron-photon interaction
that induces the transition is contact, the expression for the threshold
scattering intensity can be obtained from (\ref{xray}) in the form \cite
{sokolov}
\begin{eqnarray}
\mathcal{I}_{\omega ^{\prime }} &\propto &\sum_{\sigma \sigma ^{\prime
}}\int_0^\infty dt_1\int_0^\infty dt_2\exp \left[ -i(\omega ^{\prime
}-E_0)(t_1-t_2)-\Gamma (t_1+t_2)\right]  \nonumber \\
&&\ \langle 0|c_\sigma \exp (i\mathcal{H}_ft_1)c_{\sigma ^{\prime
}}^{\dagger }\exp [i\mathcal{H}_i(t_2-t_1)]c_{\sigma ^{\prime }}\exp (-i%
\mathcal{H}_ft_2)c_\sigma ^{\dagger }|0\rangle  \label{ffff}
\end{eqnarray}
where $H_f$ and $H_i$ are conduction-electron Hamiltonians with and without
core hole, respectively. The complicated correlation function in (\ref{ffff}%
) can be decoupled in the ladder approximation which is exact for the empty
conduction band. Then we obtain \cite{sokolov}

\begin{eqnarray}
\mathcal{I}_{\omega ^{\prime }} &\propto &W^2L(\omega ^{\prime }-E_0),
\label{zz} \\
L(E) &=&\left| \sum_\sigma G_{00}^\sigma (E+i\Gamma )\right| ^2  \nonumber
\end{eqnarray}
where $G_{00}^\sigma $ is given by (\ref{g00}), $W$ is a transition matrix
element (in a simple model of Ref.\onlinecite{sokolov} this is imaginary
part of the hole potential). Owing to a jump in the density of states at the
Fermi level, the NQP part of the Green's function contains a large logarithm
$\ln (D/[\omega ^{\prime }-E_0+i\Gamma ])$ at small $|\omega ^{\prime }-E_0|$%
, $D$ being a bandwidth. It means that the corresponding contribution to the
elastic x-ray scattering intensity ($\omega ^{\prime }=E_0$) is enhanced by
a factor of $\ln (D/\Gamma )$. Provided that the NQP contribution dominates
over the majority quasiparticle contribution (which is possible for small $%
\Gamma $ only), the enhancement factor in $\mathcal{I}_{\omega ^{\prime }}$
is $\ln ^2(D/\Gamma ).$ This makes a quantitative estimation for the
qualitative effect discussed in Ref. \onlinecite{kurm}. Of course, smearing
of the jump in the NQP density of states by spin dynamics is irrelevant
provided that $\Gamma \gtrsim \overline{\omega }$ ($\overline{\omega }$ is a
characteristic magnon frequency).

Figs. 6-7 show the function $L(E)$ for both signs of the $s-d$ exchange
parameter $I.$ One can see that NQP contributions result in a maximum at $E=0
$ for $I>0$ and in a minimum for $I>0$. The difference is due to different
signs of the NQP jumps at the Fermi level and, consequently, of the
corresponding logarithms in the real part. The numerical results are also
presented for the Hartree-Fock approximation (where the Green's functions $%
G_{\mathbf{k}}^\sigma (E)=(E-t_{\mathbf{k}\sigma })^{-1}\ $are substituted
into (\ref{g00}), (\ref{zz})) which does not take into account NQP effects.
The deviation turns out to be asymmetric because the NQP contributions are
asymmetric with respect to the Fermi level. Of course, our calculation is
not quite strict (in particular, the NQP terms in $W$ should be in principle
considered). However, this corresponds to account of main singular
contributions.

\section{Conclusions and discussion}

To conclude, we emphasize the main points of the present work. First, the
non-quasiparticle (NQP) states (in HMF these are the only minority states
which are present near the Fermi level) do manifest themselves in the core
level spectroscopy. This demonstrates an important role of correlation
effects beyond mean field (or Fermi-liquid) picture in metallic magnets (for
a general discussion, see Ref.\onlinecite{ufn}). Moreover, due to
interference of interactions of electrons with magnons and with core holes
an enhancement of NQP contributions is possible in comparison with the
initial density of states.

From a formal point of view, the results obtained in Sect.2 yield a
non-trivial analytical solution of a three-body problem. For the case $I>0$
which takes place for the most of known ``spintronic'' materials (including
colossal magnetoresistance manganites) the ``ladder'' approximation provides
probably a complete solution describing basic physics for both
semiconductors and metals. For the case $I<0$ some analogue of the Kondo
effect occurs which may lead to complicated theoretical issues. It is not
too important pragmatically since in the spectroscopy problems the
Kondo-like divergences are cut at the inverse core level lifetime $\Gamma $.
However, further theoretical investigations are of interest to obtain a more
consistent scheme.

To describe HFM state, we used in our calculations a simple model
of saturated ferromagnet (Figs.1-2). However, the numerical
results can change substantially for more realistic bandstructure
of HMF. A smooth enough band structure is expected, e.g., for such
strong itinerant ferromagnets as pyrite systems
\textrm{Fe}$_{1-x}\mathrm{Co}_x\mathrm{S}_2,$ see Refs. \cite
{pyr} (according to these electronic structure calculations,
$\mathrm{CoS}_2$ is an almost HMF system). On the other hand,
hybridization peaks near the
energy gap exist in the HMF from the class of the Heusler alloys and in CrO$%
_2$ (see Ref.\cite{ufn}). Thus calculations with the use of a concrete
electron structure of HMF are needed for a detailed comparison with x-ray
spectra, which can be done on the basis of equations obtained in the present
paper. A more simple situation is expected in some ferromagnetic
semiconductors, e.g., chalcogenide spinels HgCr$_2$Se$_4$ and CdCr$_2$Se$_4$
\cite{beb}. The conduction band in HgCr$_2$Se$_4$ has a spherical symmetry.
In the paramagnetic phase the valence band is fourfold degenerate at the
point $\Gamma ,$ and below the Curie point the equal-energy surfaces are
transformed into ellipsoids due to exchange field. In CdCr$_2$Se$_4$ the
bottom of $s$-like conduction band is at the point $\Gamma $ (symmetry $%
\Gamma _1$). Thus a parabolic spectrum model can be used directly for $n$%
-type spinels.

To probe the ``spin-polaron'' nature of the NQP states more explicitly, it
would be desirable to use spin-resolved spectroscopical methods such as
x-ray magnetic circular dichroism (XMCD, for a review see Ref.%
\onlinecite{ebert}). Owing to interference of electron-magnon scattering and
``exciton'' effects (interaction of electrons with the core hole) the NQP
contributions to x-ray spectra can be considerably enhanced in comparison
with those to DOS of the ideal crystal.

The research described was supported in part by Grant No. 02-02-16443 from
the Russian Basic Research Foundation, by the Russian Science Support
Foundation and by the Netherlands Organization for Scientific Research
(Grant NWO 047.016.005).

\section*{Appendix}

We consider the resolvent
\begin{equation}
R(E)=\sum_{\mathbf{k}}\frac{n_{\mathbf{k}}}{E-t_{\mathbf{k}}}=\int_{-1}^\mu
dE^{\prime }\frac{\rho (E^{\prime })}{E-E^{\prime }}
\end{equation}
for the semielliptic band with the width of $D=2\ $and the bare density of
states
\begin{equation}
\rho (E)=\frac 2\pi \sqrt{1-E^2},
\end{equation}
$\mu $ being the chemical potential (the Fermi energy). We have to calculate
the function
\begin{equation}
R\left( z\right) =\frac 2\pi \int\limits_{-1}^\mu \frac{dE\sqrt{1-E^2}}{z-E}%
=\frac 2\pi \int\limits_{-\mu }^1\frac{dE\sqrt{1-E^2}}{z+E}
\end{equation}
We make the substitution $E=\cos \phi ,\phi \in \left( 0,\pi /2\right) $.
Then $\mu =-\cos \phi _0$ ($\mu <0$) and
\begin{equation}
R\left( z\right) =\frac 2\pi \int\limits_0^{\phi _0}\frac{d\phi \left[
1-\left( z+\cos \phi -z\right) ^2\right] }{z+\cos \phi }=\frac 2\pi \left[
\left( 1-z^2\right) M\left( z\right) +z\phi _0-\sqrt{1-\mu ^2}\right]
\end{equation}
where
\begin{equation}
M\left( z\right) =\int\limits_0^{\phi _0}\frac{d\phi }{z+\cos \phi }
\end{equation}
On substituting $\zeta =e^{i\phi },$ $\zeta _0=e^{i\phi _0}$ we obtain $\mu
=-\frac 12\left( \zeta _0+1/\zeta _0\right) $ or
\begin{equation}
\zeta _0=-\mu +i\sqrt{1-\mu ^2}
\end{equation}
Then we have
\[
M\left( z\right) =-2i\int\limits_1^{\zeta _0.}\frac{d\zeta }{\zeta
^2+2z\zeta +1}=\frac{2i}{\zeta _1-\zeta _2}\int\limits_1^{\zeta _0.}d\zeta
\left( \frac 1{\zeta -\zeta _2}-\frac 1{\zeta -\zeta _1}\right)
\]
where
\begin{equation}
\zeta _{1,2}=-z\pm \sqrt{z^2-1}
\end{equation}
This gives us a desired complex function
\begin{equation}
M\left( z\right) =\frac i{\sqrt{z^2-1}}\left( \ln \frac{\zeta _0-\zeta _2}{%
1-\zeta _2}-\ln \frac{\zeta _0-\zeta _1}{1-\zeta _1}\right)
\end{equation}

Note that the standard choose of the branch of logarithmic function, $\ln
z=\ln |z|+i\arg (z),$ corresponds to the advanced Green's function. To
consider the retarded Green's function, we have to change the sign of the
imaginary part. On the real axis we obtain

\begin{equation}
\mathrm{Im}R(E)=\left\{
\begin{array}{cc}
-2\sqrt{1-E^2}, & -1<E<\mu  \\
0 & \mathrm{otherwise}
\end{array}
\right.
\end{equation}
\begin{equation}
\mathrm{Re}R(E)=\frac 2\pi \left[ (1-E^2)M_1(E)-\sqrt{1-\mu ^2}+E(\pi
/2+\arcsin \mu )\right]
\end{equation}
For $|E|>1$ we have
\begin{equation}
M_1(E)=\frac 2E(1-1/E^2)^{-1/2}\arctan \left[ \frac{1+u}{1-u}\left( \frac{E-1%
}{E+1}\right) ^{1/2}\right]
\end{equation}
where
\[
u=\mu /(1+\sqrt{1-\mu ^2})
\]
For $|E|<1$ we derive
\begin{equation}
M_1(E)=\frac 1{\sqrt{1-E^2}}\left[ \ln \left| \frac{Eu-1-\sqrt{1-E^2}}{Eu-1+%
\sqrt{1-E^2}}\right| +\ln \left| \frac{E+1-\sqrt{1-E^2}}{E+1+\sqrt{1-E^2}}%
\right| \right]
\end{equation}
The logarithmic terms can be smeared with a corresponding smearing of a jump
in the real part,
\begin{eqnarray}
\ln |x-a| &\rightarrow &\frac 12\ln [(x-a)^2+\delta ^2], \\
\theta (x) &=&\frac 12+\mathrm{sign}x\rightarrow \frac 12+\frac 1\pi \arctan
\frac x\delta
\end{eqnarray}
The integral
\[
\widetilde{R}(E)=\sum_{\mathbf{k}}\frac{1-n_{\mathbf{k}}}{E-t_{\mathbf{k}}}
\]
can be obtained as
\[
\widetilde{R}(E)=\left. R(E)\right| _{\mu =1}-R(E)
\]
On the real axis we have
\begin{equation}
\mathrm{Re}\left. R(E)\right| _{\mu =1}=\left\{
\begin{array}{cc}
2E, & |E|<1 \\
2[E-(E^2-1)^{1/2}] & |E|>1
\end{array}
\right.
\end{equation}

\newpage

\begin{figure}[tbp]
\includegraphics[clip]{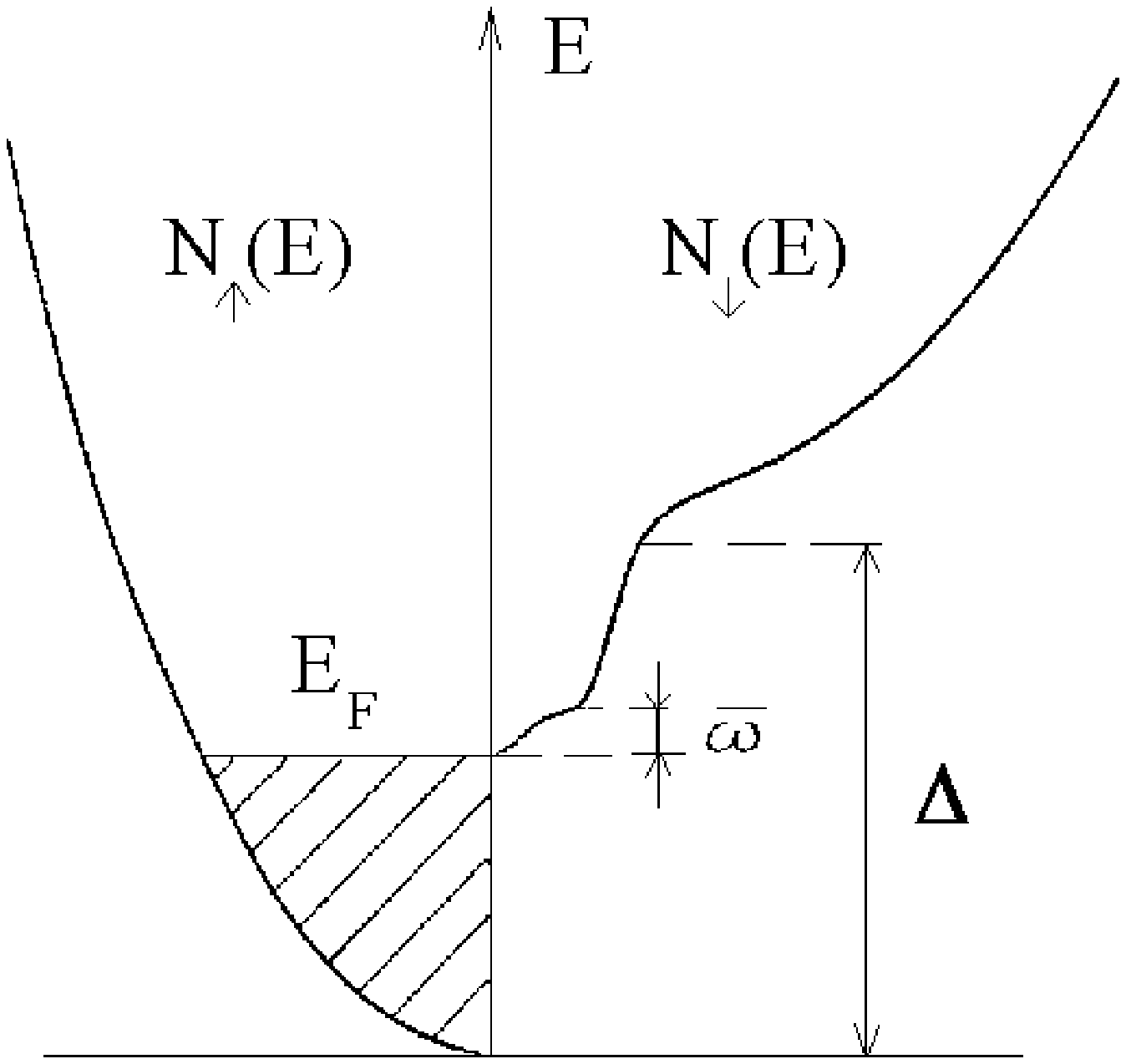}
\caption{Density of states in a half-metallic ferromagnet with $I>0$
(schematically). Non-quasiparticle states with $\sigma =\downarrow $ occur
above the Fermi level.}
\label{fig:1}
\end{figure}

\begin{figure}[tbp]
\includegraphics[clip]{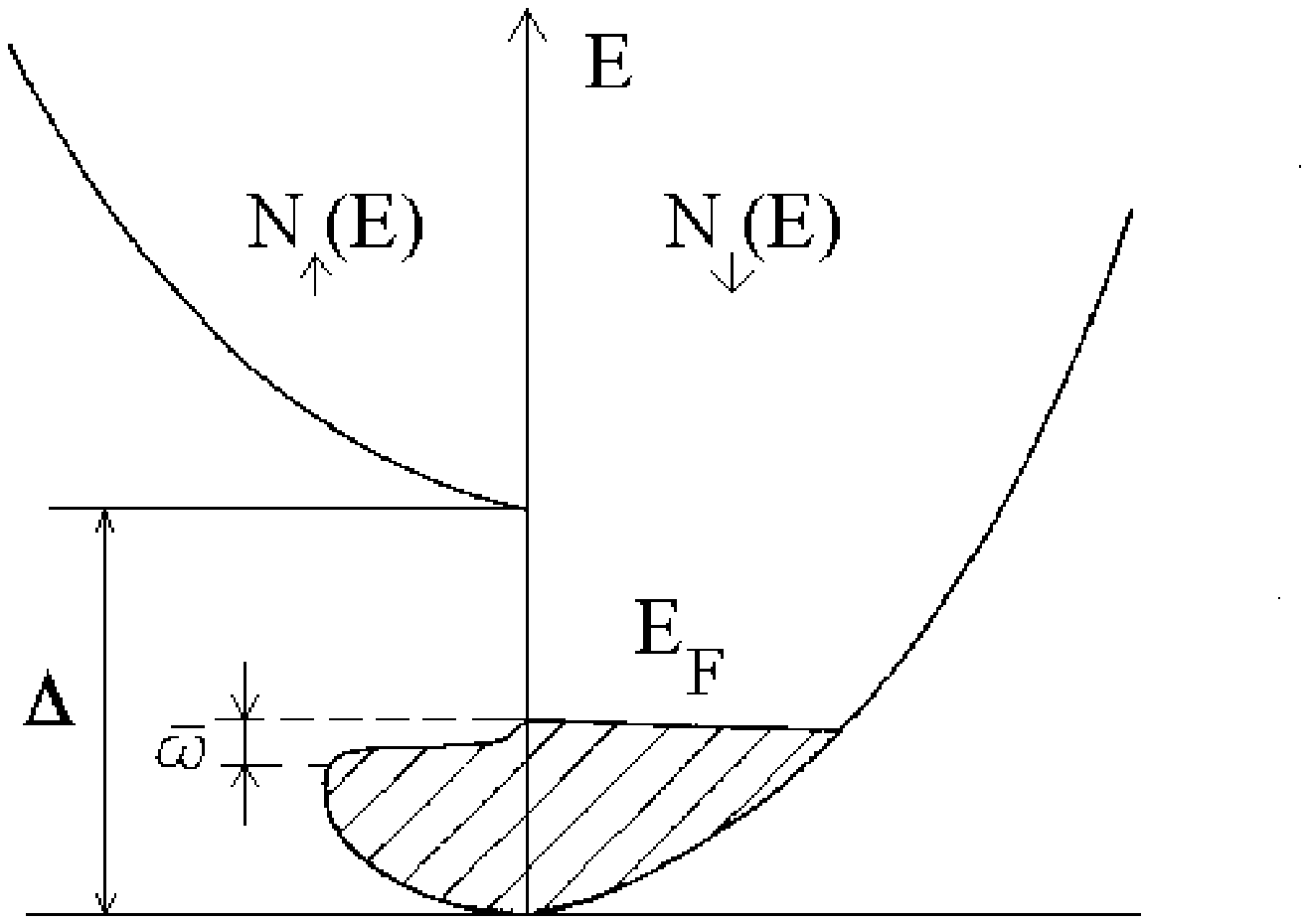}
\caption{Density of states in a half-metallic ferromagnet with $I<0$
(schematically). Non-quasiparticle states with $\sigma =\uparrow $ occur
below the Fermi level.}
\label{fig:2}
\end{figure}

\begin{figure}[tbp]
\includegraphics[clip]{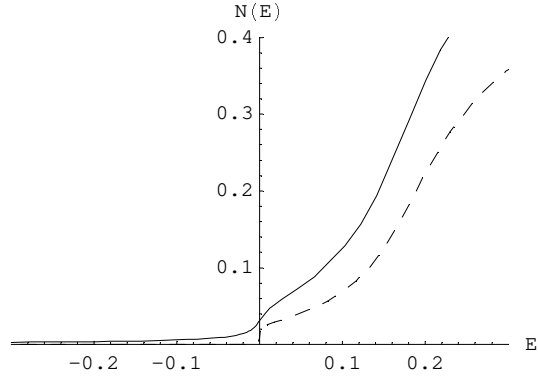}
\caption{The local density of states $N_{\mathrm{loc}}^{\downarrow }(E)$
(solid line) for a half-metallic ferromagnet with $S=1/2,I=0.3,\delta =0.01$
in the presence of the core hole potential $U=0.2.$ The dashed line shows
the DOS $N_{\downarrow }(E)$ for the ideal crystal with spin dynamics being
neglected. The value of $E_F$ calculated from the band bottom is 0.15. The
energy $E$ is referred to the Fermi level.}
\label{fig:3}
\end{figure}

\begin{figure}[tbp]
\includegraphics[clip]{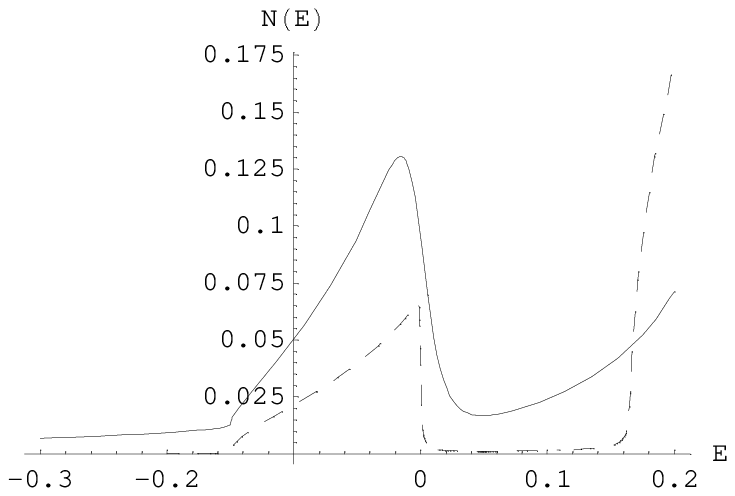}
\caption{The local density of states $N_{\mathrm{loc}}^{\uparrow }(E)$
(solid line) for a half-metallic ferromagnet with $S=1/2,I=-0.3,\delta
=0.025 $ in the presence of the core hole potential $U=0.2.$ The dashed line
shows the DOS $N_{\uparrow }(E)$ for the ideal crystal. The value of $E_F$
calculated from the band bottom is 0.15.}
\label{fig:4}
\end{figure}

\begin{figure}[tbp]
\includegraphics[clip]{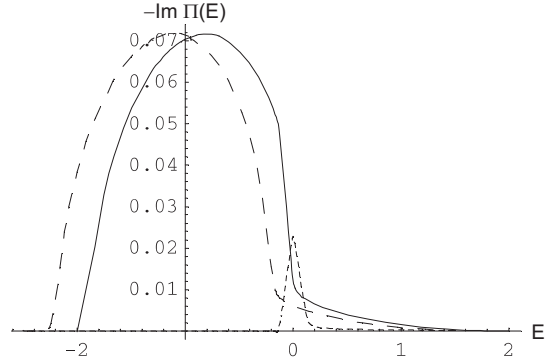}
\caption{The functions $-\mathrm{Im}\Pi _0(E)$ (solid line), $-\mathrm{Im}%
\Pi _1(E)$ (dashed line) and $-\mathrm{Im}\Pi_2(E)$ (short-dashed line) for
a half-metallic ferromagnet with $S=1/2,I=-0.3,\delta =0.01$. The value of $%
E_F $ calculated from the band bottom is 0.15.}
\label{fig:5}
\end{figure}

\begin{figure}[tbp]
\includegraphics[clip]{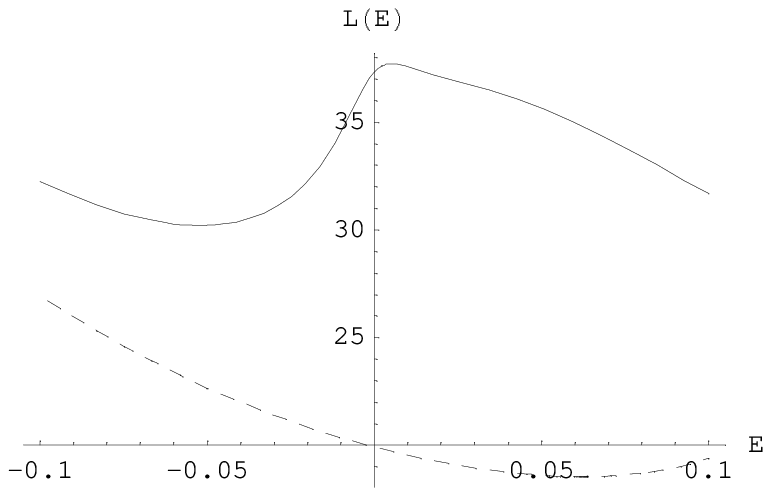}
\caption{The function $L(E)$ with account of NQP contributions (solid line),
and in the Hartree-Fock approximation (dashed line) for a half-metallic
ferromagnet with $S=1/2,I=0.3,\Gamma =0.01$. The value of $E_F $ calculated
from the band bottom is 0.15, the hole potential is $U=0.3$.}
\label{fig:6}
\end{figure}

\begin{figure}[tbp]
\includegraphics[clip]{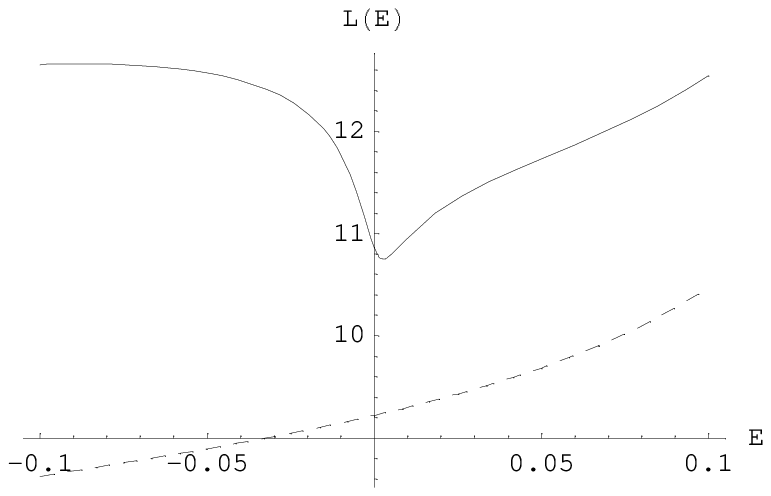}
\caption{The function $L(E)$ with account of NQP contributions (solid line),
and in the Hartree-Fock approximation (dashed line) for a half-metallic
ferromagnet with $S=1/2,I=-0.3,\Gamma =0.01$. The value of $E_F $ calculated
from the band bottom is 0.15, the hole potential is $U=0.1$.}
\label{fig:7}
\end{figure}

\end{document}